# A framework to identify supercritical and subcritical Turing bifurcations: Case study of a system sustaining cubic and quadratic autocatalysis

Deepak Kumar[1], Uttam Kumar[1] and Subramaniam Pushpavanam [1,*]

[1]Department of Chemical Engineering, Indian Institute of Technology Madras, Chennai 600036

*Corresponding author: Email address spush@iitm.ac.in

## Abstract

In this work, we focus on an autocatalytic reaction-diffusion model and carry out multiple scale weakly nonlinear analysis. A cubic and a quadratic autocatalytic reaction system is analysed. We develop a framework to identify the critical surfaces in parameter space across which the nature of the Turing bifurcation changes from supercritical to subcritical. These are verified by direct numerical simulations of the system. Using weakly nonlinear analysis, we derive equations up to the fifth order that governs the amplitude of the spatial patterns. The limit point of the bifurcating solution is captured accurately by extending the analysis to the fifth order for the case of subcritical bifurcation. The numerical solutions are in good agreement with the predictions of the weakly nonlinear analysis for supercritical bifurcations. We show that when multiple steady states arise Turing patterns can coexist with another spatially uniform steady states. Furthermore, we show that our framework can be extended to get different patterns like squares and hexagons in a two-dimensional domain. We show that the shape of Turing patterns is influenced by the domain size. This shows that the geometry can influence the kind of patterns formed in natural systems. This study will aid the experimentalist identify operating conditions where Turing patterns can be obtained.

## 1. Introduction

Various natural systems exhibit distinct self-regulated patterns due to the presence of nonlinear interactions among species. These include patterns on shells, during embryogenesis [1], in RNA synthesis [2]. They also arise in the geosciences [3], and spatio-temporal dynamics in predator-prey models [4]. Systems which sustain nonlinear chemical kinetics with diffusion exhibit spatiotemporal patterns [5]. The kinetics are typically characterized by positive feedback effects [6–8]. Such behavior of a self-regulated system is modelled by a system of reaction-diffusion equations [9,10]. The formation of spatiotemporal patterns requires at least two species. Gierer and Meinhardt have analysed a system where one species functions as an activator and the other as an inhibitor which exhibits steady spatial patterns [6]. Reaction-diffusion equations has been widely used in study of population biology and epidemiology [11–13]. Kondo and Miura, in their review [8], discussed how reaction-diffusion models can predict pattern formation in biological systems. Andrew et al. studied Turing instabilities for spatially heterogeneous reaction-diffusion system. They found that the structure of unstable modes differs substantially from typical trigonometric functions [14].

Weakly nonlinear analysis has been used to study complex dynamical systems in various fields of science and engineering [15]. Weakly nonlinear analysis was performed on Rayleigh-Benard convection of a fluid layer bounded between poorly conducting boundaries to analyse the solutions that depict two-dimensional roll patterns [16]. Other hydrodynamic instabilities like Rayleigh Taylor instability, Kelvin-Helmholtz instability and double diffusive instability have been analysed using this approach [17–19]. In the recent years, weakly nonlinear analysis helped to describe spatial pattern formation in electrodeposition [20,21]. The bifurcation behavior of a prototype reaction diffusion model was investigated [22]. In their work, the challenges in scaling reaction-diffusion equations and deriving the asymptotically valid solutions were discussed. The classical Stuart-Landau equation up to the third order cannot capture the stable amplitude branch in a subcritical bifurcation [23,24]. Consequently, a quintic Stuart Landau equation is required to obtain the stable branch of the bifurcating solution. A reaction-diffusion system with a nonlinear diffusion term was studied in [25] and amplitude equations for both Hopf and Turing instability were obtained. Consolo et.al [26] considered a reaction-advection-diffusion model to study the Turing patterns whose amplitude varies in space and time using Ginzburg-Landau equation. The amplitudes of the spatiotemporal patterns emerging in a predator prey model near the Turing and Turing-Hopf bifurcation point were characterized using amplitude equations [27].

The combined cubic and quadratic autocatalytic reactions are commonly encountered in asymmetric autocatalysis. Here, the presence of an enantiomer influences the replication of the substrate into the enantiomer [28,29]. Here, the enantiomer $(E)$ stimulates the substrate $(S)$ to form that enantiomer following the reaction

$$S + E \rightarrow 2E.$$

Similar systems arise in the inter conversion of optical isomers $(E')$ into $(E)$ and vice versa, given by

$$E' + E \rightarrow 2E,$$

$$E + E' \rightarrow 2E'.$$

This kind of autocatalytic reaction system was previously studied by our group [30,31]. In those studies, the steady state behaviour was analysed and the conditions for complete conversion of one of the species was obtained. Uttam and Pushpavanam analysed an autocatalytic reaction diffusion system and studied the effect of subdiffusion on the Turing instability using linear stability analysis [31]. In this work, we analyse the same system using a weakly nonlinear analysis for completeness and presentation clarity. This approach helps us to choose parameters for supercritical and subcritical bifurcation for asymmetric autocatalysis reaction diffusion. Typically, nonlinear systems are characterized by several parameters. It is hence a challenge to identify regions in parameter space where different instabilities arise. This work focuses on developing a framework to determine the parameter values under which the system exhibits Turing patterns. Critical surfaces are identified across which the nature of the bifurcation changes from supercritical to subcritical. We also show that the bifurcation of a Turing pattern from a steady state can coexist with a spatially uniform steady state. We also show how the approach can be used to identify different 2D patterns such as squares and hexagons in nonlinear systems.

We have structured the article as follows. The governing equations of the model are discussed in Section 2. Section 3 focuses on deriving the conditions for the onset of Turing patterns and investigates the critical surface where the bifurcation nature transitions from supercritical to subcritical. We derive the Stuart-Landau amplitude equation up to the quintic order which aids in determining the limit point of the bifurcating branch for subcritical bifurcation. The framework used in this work is detailed and the key results are presented here. In Section 4, we conclude the paper by summarizing the main findings and contributions of our work.

## 2. Model description and governing equations

The system under investigation has two species $A$ and $B$ undergoing two autocatalytic reactions. The autocatalytic generation of $A$ and $B$ follow cubic and quadratic kinetics respectively. Additionally, we consider species $B$ can be produced by species $A$ via a non-autocatalytic step. These reactions are given by

$$2A + B \rightarrow 3A, \quad r_a = k_{ga}A^2B, \tag{1a}$$

$$A + B \rightarrow 2B, \quad r_b = k_{gb}AB, \tag{1b}$$

$$A \rightarrow B, \quad r_b = k_{ab}A. \tag{1c}$$

Furthermore, reactants $A$ and $B$ follow first order kinetics and decay to produce inert species $I_1$ and $I_2$, respectively.

$$A \rightarrow I_1, \quad -r_a = k_{da}A, \tag{2a}$$

$$B \rightarrow I_2, \quad -r_b = k_{db}B. \tag{2b}$$

Species $A$ and $B$ are produced from precursors $P_1, P_2$ with rate constants $k_{pa}$ and $k_{pb}$ respectively. The precursors are assumed to be in excess and hence their concentrations are taken constant.

$$P_1 \rightarrow A, \quad r_a = k_{pa}P_1, \tag{3a}$$

$$P_2 \rightarrow B, \qquad r_b = k_{pb}P_2. \tag{3b}$$

The spatio-temporal evolution of concentration of species $A$ and $B$ with time ($\tau$) and space ($X$) is governed by

$$\frac{dA}{d\tau} = D_a \nabla^2 A + k_{pa}P_1 + k_{ga}A^2B - k_{gb}AB - k_{ab}A - k_{da}A, \tag{4a}$$

$$\frac{dB}{d\tau} = D_b \nabla^2 B + k_{pb}P_2 - k_{ga}A^2B + k_{gb}AB + k_{ab}A - k_{db}B. \tag{4b}$$

Here, $D_i$ represents the diffusivity of the species $i$.

We define the following dimensionless variables:

$$a = \left(\frac{k_{ga}}{k_{da}}\right)^{\frac{1}{2}} A, b = \left(\frac{k_{ga}}{k_{da}}\right)^{\frac{1}{2}} B, P_a = \frac{k_{pa}}{k_{da}}\left(\frac{k_{ga}}{k_{da}}\right)^{\frac{1}{2}} P_1, P_b = \frac{k_{pb}}{k_{da}}\left(\frac{k_{ga}}{k_{da}}\right)^{\frac{1}{2}} P_2,$$

$$g_b = k_{gb}(k_{da}k_{ga})^{-\frac{1}{2}}, K_a = \frac{k_{ab}}{k_{da}}, d_b = \frac{k_{db}}{k_{da}}, t = k_{da}\tau, d = \frac{D_a}{D_b}, x = \left(\frac{k_{da}}{D_b}\right)^{\frac{1}{2}} X.$$

The evolution of concentrations of the two species for a well-mixed system is governed by

$$\frac{da}{dt} = P_a + a^2 b - g_b ab - K_a a - a = F(a,b), \tag{5a}$$

$$\frac{db}{dt} = P_b - a^2 b + g_b ab + K_a a - d_b b = G(a,b). \tag{5b}$$

Including diffusion results in the following set of reaction diffusion equations which govern the spatio-temporal evolution of the two species

$$\frac{\partial a}{\partial t} = d\nabla^2 a + F(a,b), \tag{6a}$$

$$\frac{\partial b}{\partial t} = \nabla^2 b + G(a,b). \tag{6b}$$

Here, $\nabla^2 \equiv \frac{\partial^2}{\partial x^2} + \frac{\partial^2}{\partial y^2}$ and the parameter $d$ represents the ratio of the diffusivity of species $A$ to that of species $B$.

## 3. Results and discussion

At steady state, the rate of change of concentration is zero. The steady-state solutions for $a_{ss}$ and $b_{ss}$ for a spatially uniform state are given by roots of

$$f(a_{ss}) = -\frac{a_{ss}^3}{d_b} + a_{ss}^2\left(\frac{g_b}{d_b} + \frac{P_a}{d_b} + \frac{P_b}{d_b}\right) - a_{ss}\left(1 + K_a + \frac{g_b P_a}{d_b} + \frac{g_b P_b}{d_b}\right) + P_a = 0, \qquad (7a)$$

$$\begin{aligned}g(b_{ss}) = &-b_{ss}^3 d_b^2 + b_{ss}^2(2d_b P_a + 2d_b P_b - d_b g_b) \\ &+ b_{ss}(g_b P_a - P_a^2 + g_b P_b - 2P_a P_b - P_b^2 - d_b - d_b K_a) + K_a P_a + P_b \\ &+ K_a P_b = 0.\end{aligned} \qquad (7b)$$

The steady states of the well mixed system are governed by cubic polynomials (7a-7b). Hence, the system can exhibit a maximum of three steady states for some parameters.

Our interest lies in determining conditions when the system exhibits super and subcritical bifurcations that give rise to Turing patterns. We first establish a framework to determine the conditions when the system exhibits 1D Turing patterns. To begin with, we analyze the system when it has only one steady state. We then focus on the system when three steady states co-exist, and Turing patterns arise when one of the stable steady states becomes unstable. Finally, we discuss the formation of 2D Turing patterns.

## 3.1. 1-Dimensional patterns

### 3.1.1. Single steady state

**Fig.1**(a) and 1(b) depict bifurcation diagrams for species $A$ and $B$ respectively. They depict how the steady state concentrations of the two species vary with the bifurcation parameter $P_a$. The system has three steady states for $P_a$ lying in the range of (0.2886 – 0.375). It also exhibits a Hopf bifurcation at $P_a = 1.1413$. For this set of parameters, this system exhibits both static and dynamic instability.

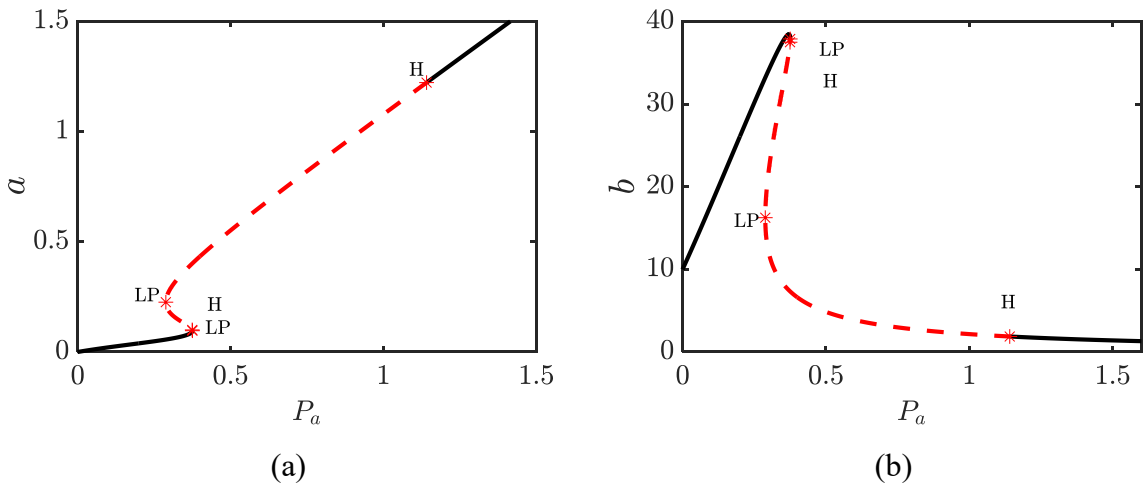

(a)            (b)

**Fig.1** The bifurcation diagram of the species $A$ and $B$ with the bifurcation parameter $P_a$ for $g_b = 0.12, P_b = 0.10, d_b = 0.01$ and $K_a = 2$. The solid line represents the stable steady state while the dashed line represents the unstable steady state.

## Linear stability analysis of steady state

The stability of the steady state is governed by the eigenvalues of the Jacobian matrix ($J$) computed at the equilibrium point $(a_{ss}, b_{ss})$.

The Jacobian matrix ($J$) is

$$J = \begin{pmatrix} F_a & F_b \\ G_a & G_b \end{pmatrix}. \tag{8}$$

Here $F_i$ and $G_i$ denote the derivatives of $F$ and $G$ with respect to concentration of species $i$ computed at the steady state. In this section, we investigate the behavior of the system where only one steady state is present. For $P_a = 1.2$, the system has a unique stable steady state $(1.282, 1.77)$.

The stability conditions for a steady state are

$$Det(J) = F_a G_b - F_b G_a > 0 \text{ and } Tr(J) = F_a + G_b < 0. \tag{9}$$

We introduce diffusion in the system and check the stability of the same steady state in the reaction diffusion system. For this we impose infinitesimally small perturbation of the form $e^{s_1 t} e^{ikx}$. Here, $s_1$ denotes the growth rate of perturbation and $k$ is the wavenumber of the disturbance. We substitute $a = a_{ss} + e^{s_1 t} e^{ikx} a_1$ and $b = b_{ss} + e^{s_1 t} e^{ikx} b_1$ into equations (6a) and (6b). The linearized equations which govern the perturbations are given by

$$\frac{da_1}{dt} = F_a a_1 + F_b b_1 - dk^2 a_1, \tag{10a}$$

$$\frac{db_1}{dt} = G_a a_1 + G_b b_1 - k^2 b_1. \tag{10b}$$

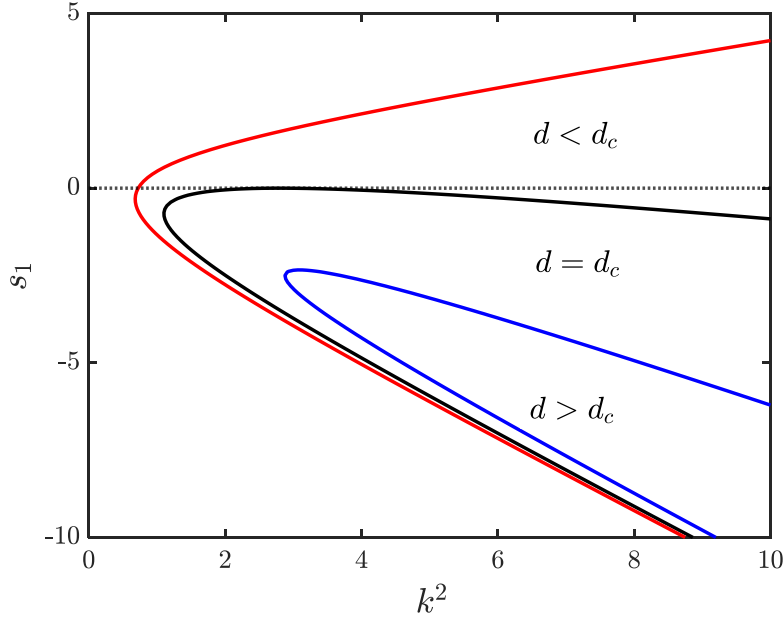

**Fig.2** Dispersion curve showing the growth rate of the perturbation of the $k$th mode for $P_a = 1.2, g_b = 0.12, P_b = 0.1, d_b = 0.01, K_a = 2$ and $d_c = 0.1893$.

The stability of the steady states for the reaction diffusion system is governed by the eigenvalues of the Jacobian matrix ($\mathbf{J}_{RD}$)

$$\mathbf{J}_{RD} = \begin{pmatrix} F_a - k^2 d & F_b \\ G_a & G_b - k^2 \end{pmatrix}. \tag{11}$$

The mathematical relationship between the growth rate of perturbations and their spatial scale, defined by the wave number is the dispersion relationship. This is given by

$$s_1^2 - Tr(\mathbf{J}_{RD})s_1 + Det(\mathbf{J}_{RD}) = 0, \tag{12}$$

where,

$$Tr(\mathbf{J}_{RD}) = Tr(\mathbf{J}) - k^2(1 + d), \tag{13}$$

$$Det(\mathbf{J}_{RD}) = k^4 d - k^2(F_a + dG_b) + Det(\mathbf{J}). \tag{14}$$

From equation (13), $Tr(\mathbf{J}_{RD})$ remains negative even after including diffusion. Hence an instability cannot occur by violation of this condition if the $Tr(\mathbf{J}) < 0$. Based on equation (14), the condition for instability requires $F_a + dG_b > 0$. Taking this into consideration, along with the condition $Tr(\mathbf{J}_{RD}) < 0$, we deduce that $F_a$ and $G_b$ should have opposite signs. In our specific system, $F_a > 0$ and $G_b < 0$. Consequently, species $A$ functions as an activator, while species $B$ acts as an inhibitor.

The surface across which the Turing patterns emerge is defined by $Det(\mathbf{J}_{RD}) = 0$ which is obtained by setting $s_1 = 0$ in equation (12). The minimum of $Det(\mathbf{J}_{RD}) = 0$ occurs at $k = k_c$ given in equation (15). By substituting this value into $Det(\mathbf{J}_{RD}) = 0$, we determine the critical value of $d$ across which the instability occurs (given by equation (16)). The expressions for $k_c$ and $d_c$ are given by

$$k_c = \left(\frac{d_c G_b + F_a}{2\, d_c}\right)^{\frac{1}{2}}, \qquad (15)$$

$$d_c = \frac{F_a G_b - 2\, F_b G_a - 2\left(F_b^2 G_a^2 - F_a G_a F_b G_b\right)^{\frac{1}{2}}}{G_b^2}. \qquad (16)$$

The growth rate of perturbations can be positive or negative depending on the value of control parameter. Fig.2 illustrates the relationship between the growth rate ($s_1$) and wave number ($k^2$) for different values of $d$. It is observed that the growth rate becomes positive when $d < d_c$, indicating that diffusion can destabilize the steady state. Conversely, for $d > d_c$, the growth rate remains negative for all wave numbers indicating a stable steady state. Here, $d_c$ is the critical value of the diffusivity ratio. The positive value of eigenvalue ($s_1$) for $d < d_c$ indicates that the system exhibits a periodic spatial pattern for sufficiently low values of $d$.

*Weakly nonlinear (WNL) stability analysis*

The linear stability analysis establishes the occurrence of Turing patterns for a set of parameter values. The nature of the bifurcation branch and its direction can be obtained using weakly nonlinear (WNL) analysis. We use multiple scale analysis [32] to rescale time and the control parameter $d$ to capture the slow dynamics of the pattern near the critical value ($d_c$). The rescaling is expressed in terms of a small parameter $\epsilon$, which is associated with the magnitude of the perturbation.

$$\frac{\partial}{\partial t} = \epsilon \frac{\partial}{\partial T_1} + \epsilon^2 \frac{\partial}{\partial T_2} + \epsilon^3 \frac{\partial}{\partial T_3} + \epsilon^4 \frac{\partial}{\partial T_4} + \epsilon^5 \frac{\partial}{\partial T_5} + O(\epsilon^6), \qquad (17a)$$

$$d = d_c + \epsilon d_1 + \epsilon^2 d_2 + \epsilon^3 d_3 + \epsilon^4 d_4 + \epsilon^5 d_5 + O(\epsilon^6). \qquad (17b)$$

The perturbed variables $a$ and $b$ are given vectorially as

$$\mathbf{u} = \mathbf{u}_{ss} + \epsilon \mathbf{u}_1 + \epsilon^2 \mathbf{u}_2 + \epsilon^3 \mathbf{u}_3 + \epsilon^4 \mathbf{u}_4 + \epsilon^5 \mathbf{u}_5 + O(\epsilon^6), \qquad (18)$$

where, $\mathbf{u}_i = \begin{pmatrix} a_i \\ b_i \end{pmatrix}$.

Collecting terms at different orders of $\epsilon$, we obtain

$O(\epsilon)$: $\qquad \mathbf{L}_c \mathbf{u}_1 = 0,$ (19a)

$O(\epsilon^2)$: $\qquad \mathbf{L}_c \mathbf{u}_2 = Q,$ (19b)

$O(\epsilon^3)$: $\qquad \mathbf{L}_c \mathbf{u}_3 = R,$ (19c)

$O(\epsilon^4)$: $\qquad \mathbf{L}_c \mathbf{u}_4 = S,$ (19d)

$O(\epsilon^5)$: $$\mathbf{L_c}\mathbf{u_5} = \mathbf{V},\quad (19e)$$

where, $\mathbf{L_c} = \mathbf{J} + \mathbf{D_c}\nabla^2$ and $\mathbf{D_c} = \begin{pmatrix} d_c & 0 \\ 0 & 1 \end{pmatrix}$.

The explicit expressions for $\mathbf{Q}, \mathbf{R}, \mathbf{S}, \mathbf{V}$ are given in supplementary material Appendix A (see equations A.2). The solution to the linear homogeneous ordinary differential equation (19a), for homogeneous Neumann boundary condition is

$$\mathbf{u_1} = A_p(T)\,\mathbf{r}\cos(k_c x), \text{ where } \mathbf{r} = \begin{pmatrix} 1 \\ M \end{pmatrix} \text{ and } M = \frac{G_a}{k_c^2 - G_b} \quad (20)$$

with $\mathbf{r} \in Ker(\mathbf{J} - k_c^2 \mathbf{D_c})$.

The terms $d_1$ and $T_1$ are responsible for resonance. Therefore, we eliminate them by just imposing $T_1 = 0$ and $d_1 = 0$. This results in the solution of the non-homogeneous differential equation

$$\mathbf{u_2} = \begin{pmatrix} a_2 \\ b_2 \end{pmatrix} = A_p^2 \begin{pmatrix} A_3 \\ A_4 \end{pmatrix}\cos(2k_c x) + \begin{pmatrix} p_3 \\ p_4 \end{pmatrix}. \quad (21)$$

The coefficients $A_3, A_4, p_3, p_4$ are obtained by equating the coefficients of different powers of $A$ from LHS and RHS in equation (19b). The explicit expressions for these coefficients are given in supplementary material Appendix A (see equations A.3). On substituting the value of $\mathbf{u_1}$ and $\mathbf{u_2}$ into the equation (19c), we obtain $\mathbf{R}$ as

$$\mathbf{R} = \left(\mathbf{r}\frac{dA_p}{dT_2} + \mathbf{R_{11}}A_p + \mathbf{R_{13}}A_p^3\right)\cos(k_c x) + \mathbf{R_{33}}\,A_p^3 \cos(3\,k_c x). \quad (22)$$

The expressions for $\mathbf{R_{11}}, \mathbf{R_{13}}, \mathbf{R_{33}}$ have been given in supplementary material Appendix A (see equations A.4). The Fredholm solvability condition is employed on system (22) to determine the value of the unknown term $\frac{dA_p}{dT_2}$. This yields the Stuart-Landau equation for amplitude $A_p(t)$

$$\frac{dA_p}{dt} = \sigma_1 A_p - L A_p^3. \quad (23)$$

The detailed procedure to find $\sigma_1$ and $L$ along with their expressions have been given in supplementary material Appendix A (see equations A.5-A.8). In the region where patterns are formed, $\sigma_1$ is always positive. Therefore, the nature of the bifurcation is dependent only on the sign of the Landau constant $L$. A positive value of $L$ corresponds to a supercritical bifurcation, whereas negative value of $L$ indicates a subcritical bifurcation. Hence, to determine the nature of the bifurcation we determine the critical surface along which $L = 0$ in the parameter space $(K_a - d_b)$. Across this, the nature of the bifurcation which gives rise to Turing patterns changes.

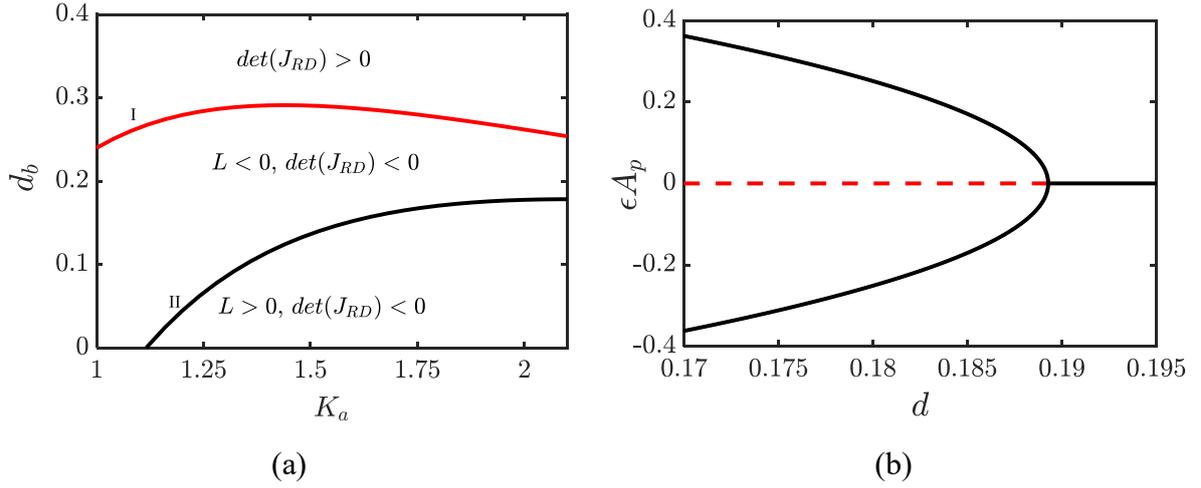

(a)                          (b)

**Fig.3** (a)The critical surface in $K_a - d_b$ space for supercritical and subcritical region for the parameters $g_b = 0.12, P_b = 0.1$ and $P_a = 1.2$. (b) The bifurcation diagram showing supercritical bifurcation for the parameters $g_b = 0.12, d_b = 0.01, P_b = 0.1, P_a = 1.2, d = 0.185, K_a = 2$ and $\epsilon = 0.1$. The solid lines represent the stable branches while the dashed line represents the unstable branch.

Curve I in Fig.3(a) represents the condition $Det(\mathbf{J_{RD}}) = 0$. This divides the entire $(K_a - d_b)$ space into two regions such that above the curve I, $Det(\mathbf{J_{RD}}) > 0$ and below it $Det(\mathbf{J_{RD}}) < 0$. Bifurcation giving rise to Turing patterns can be seen only below curve I [33]. Curve II is obtained numerically by equating $L = 0$. The nature of the bifurcation is subcritical above curve II and is supercritical below it.

*Case (1) Supercritical case:* $L > 0$

A typical bifurcation diagram for parameters in the supercritical region below curve II where $L > 0$ is shown in Fig.3(b). The system is unstable to small perturbations for $d < d_c$ and it reaches a new stable steady state whose amplitude is given by $\pm\sqrt{\sigma_1/L}$. The spatially uniform steady state is always stable for $d > d_c$. We also perform numerical simulations of the nonlinear system given by 6(a) and 6(b) to validate the solution obtained by WNL analysis. We compare the amplitude of spatial variation of species concentration obtained from numerical solutions with that predicted by the WNL analysis in Fig.4. We see a good agreement between these two solutions. The match between the two solutions is valid only for low values of $\epsilon$ i.e., close to the bifurcation point. The new steady state of the system is steady and periodic in space.

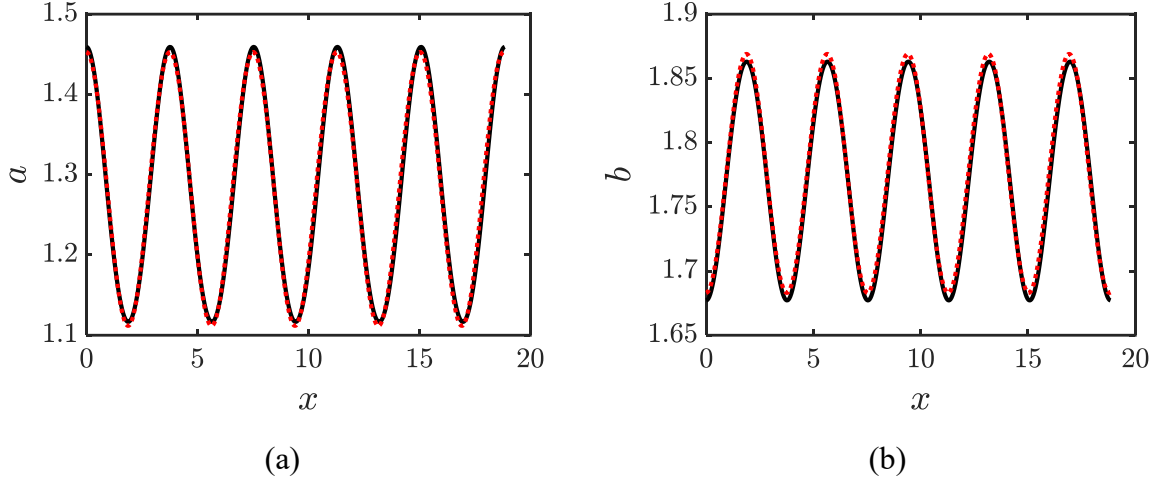

(a) (b)

**Fig.4** Comparison between the numerical solution and the weakly nonlinear analysis for the parameters $g_b = 0.12$, $d_b = 0.01$, $P_b = 0.1$, $P_a = 1.2$, $K_a = 2$, $d = 0.185$ and $\epsilon = 0.1$. The solid black line represents the numerical solution while the dotted red line represents the weakly nonlinear approximated solution.

*Case (2) Subcritical case:* $L < 0$

For subcritical bifurcation the parameters are chosen above curve II where $L < 0$. For $d > d_c$, the uniform steady state is stable. However, the system shows Turing patterns for a finite amplitude disturbance. Here, the spatially uniform steady state is stable to small disturbances but unstable to finite amplitude disturbances as shown in Fig. 5 (a). Fig. 5(a) is obtained from the WNL analysis at $O(\epsilon^3)$. At this order, the Stuart-Landau equation (23) does not predict any stable Turing branch. We extend the analysis to $O(\epsilon^5)$ to determine the stable Turing pattern branch and the turning point of the bifurcating branch. Following the procedure detailed in [34], we obtain the quintic Stuart-Landau equation

$$\frac{dA_p}{dt} = \bar{\sigma}_1 A_p - \bar{L} A_p^3 + \bar{M} A_p^5. \tag{24}$$

The derivation of this quintic amplitude equation and the explicit expressions of $\bar{\sigma}_1$, $\bar{L}$ and $\bar{M}$ are detailed in supplementary material Appendix A (see equation A.23). The quintic amplitude equation obtained admits two symmetric real stable equilibria of spatially varying solutions (denoted by solid line in Fig.5(b)). The turning point value $(d_T)$ obtained using weakly nonlinear approximation is 0.0192 as shown in Fig.5(b). The spatially periodic variation obtained numerically for subcritical bifurcation is shown in Figs. 6(a) and 6(b). The numerical solution reveal that the turning point value is 0.01962. This shows that the turning point or limit point obtained from the nonlinear analysis matches closely with that obtained using the numerical simulations.

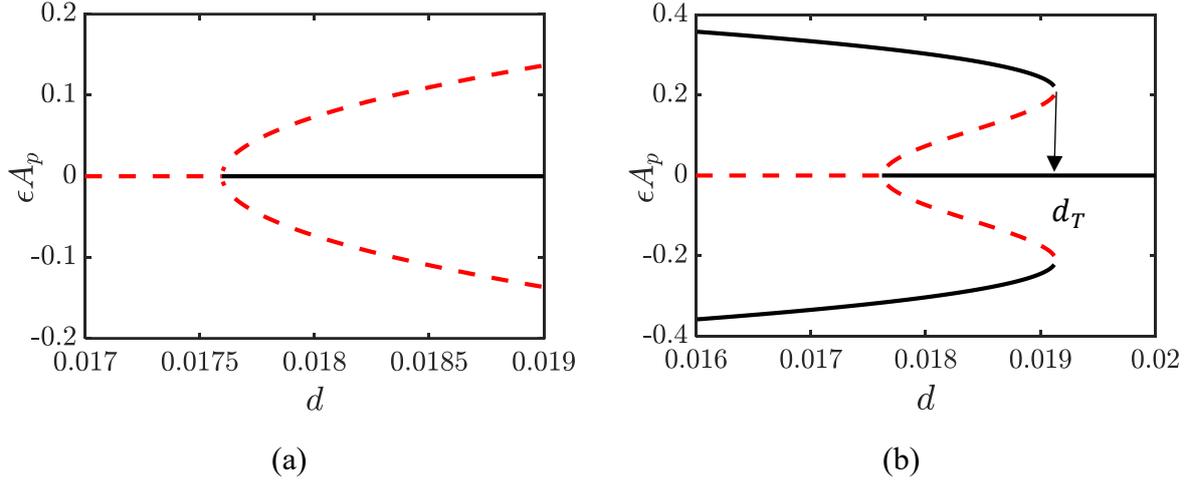

(a)             (b)

**Fig.5** The bifurcation diagram showing subcritical bifurcation for the parameters $g_b = 0.12, P_b = 0.1, d_b = 0.15, K_a = 1.3$ and $P_a = 1.2$. The solid line represents the stable branches while the dashed line represents the unstable branches for (a) Cubic Stuart – Landau equation (b) Quintic Stuart Landau equation.

The spatial variation of the Turing patterns obtained using weakly nonlinear and numerical simulations are shown in Figs.6(a) and 6(b). The solid black line represents the numerical solution and the dotted red line represents the weakly nonlinear approximated solution. The mismatch is due to the assumption in weakly nonlinear approximation of small amplitude of oscillation and slow variation of solution with respect to time and space. The numerical simulation for subcritical case (see Fig.6a) shows that the variation of species is very rapid in space and thus the rescaling of the weakly nonlinear approximation cannot capture the large amplitude of the pattern.

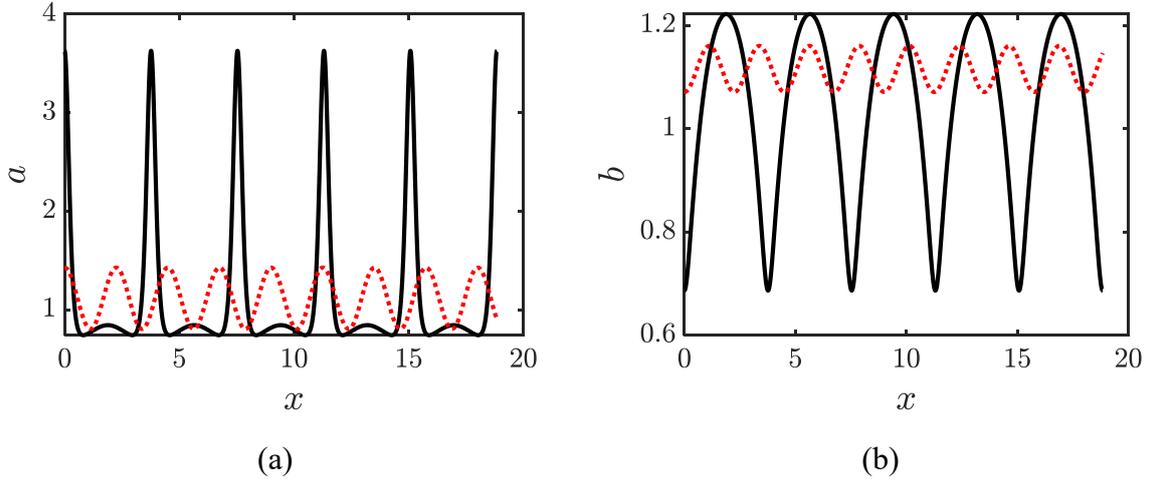

(a)             (b)

**Fig.6** Comparison between the numerical solution of the full system and the weakly nonlinear analysis for the parameters $g_b = 0.12$, $d_b = 0.15$, $P_b = 0.1$, $P_a = 1.2$, $K_a = 1.3$, $d = 0.185$ and $\epsilon = 0.1$. The black solid line represents the numerical solution while the red dotted line represents the weakly nonlinear approximated solution.

### 3.1.2. Multiple steady state

In this section, we explore the bifurcation of a Turing pattern branch from a steady state, when it coexists with other spatially uniform steady states of the system. This scenario arises in regions where multiple steady states are present, and we aim to understand the interaction between the stable Turing pattern branch with the other stable steady state. The bifurcation diagrams for species *A* and *B* are shown in Figs.7(a) and 7(b). The system has three steady states for $P_a$ lying in the range of (0.4499 − 1.8345). We select a value of the bifurcation parameter $P_a = 1.5$, for which there are three steady-states $P_1 = (2.7845, 1.07738)$, $P_2 = (0.5, 12.5)$, and $P_3 = (0.2155, 13.923)$. In the absence of diffusion, $P_1$ and $P_3$ are stable but $P_2$ is unstable.

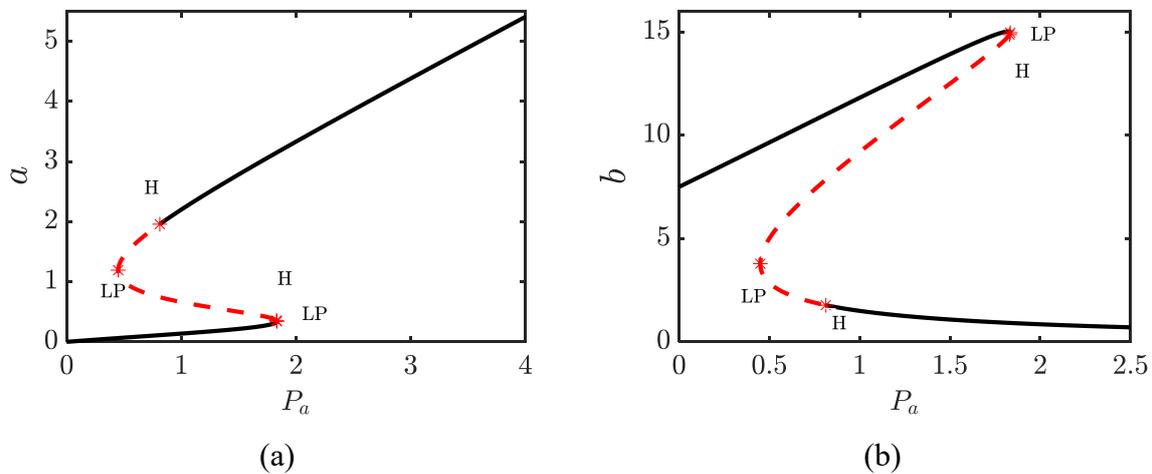

(a)          (b)

**Fig.7** The bifurcation diagram of the species concentration $a$ and $b$ on the bifurcation parameter $P_a$ for parameters $g_b = 0.5, P_b = 1.5, d_b = 0.2$, and $K_a = 2$. The solid line represents the stable steady state while the dashed line represents the unstable steady state.

The phase plane portraits of the well mixed system is depicted in Fig.8. The steady states have been represented as $(P_1, P_2, P_3)$. The equilibrium points $P_1$ and $P_3$ act as attractors and their basin of attraction is shown in the Fig.8. The point $P_2$ acts as a repeller as it is unstable and trajectories starting close to this point diverge from it. The line $a = 0.5$ acts as a separatrix. Initial conditions on the left of this boundary converges at $P_3$ while any initial condition on the right of this boundary converges at $P_1$.

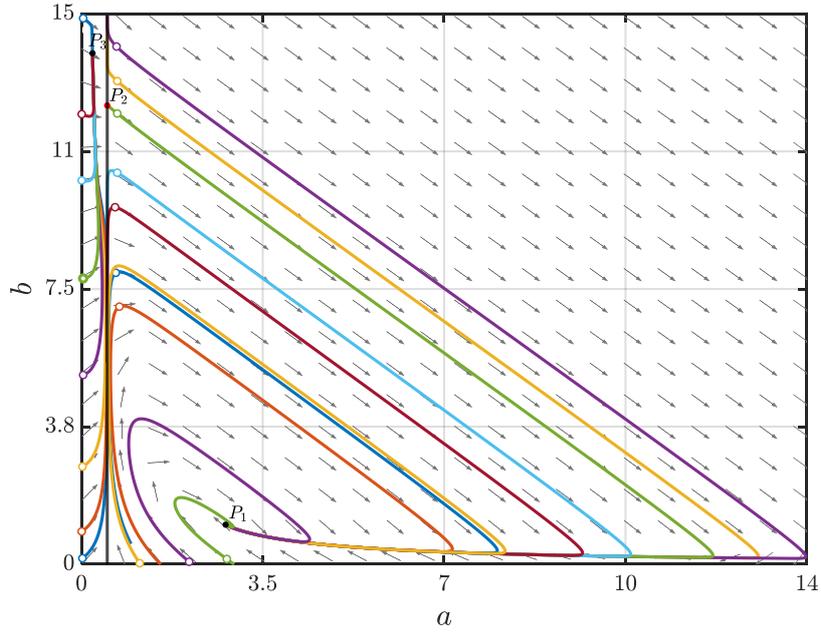

**Fig.8** The phase plane portrait of well mixed system. $P_1, P_2, P_3$ are three competitive steady states for parameter $g_b = 0.5, P_b = 1.5, d_b = 0.2, K_a = 2$ and $P_a = 1.5$.

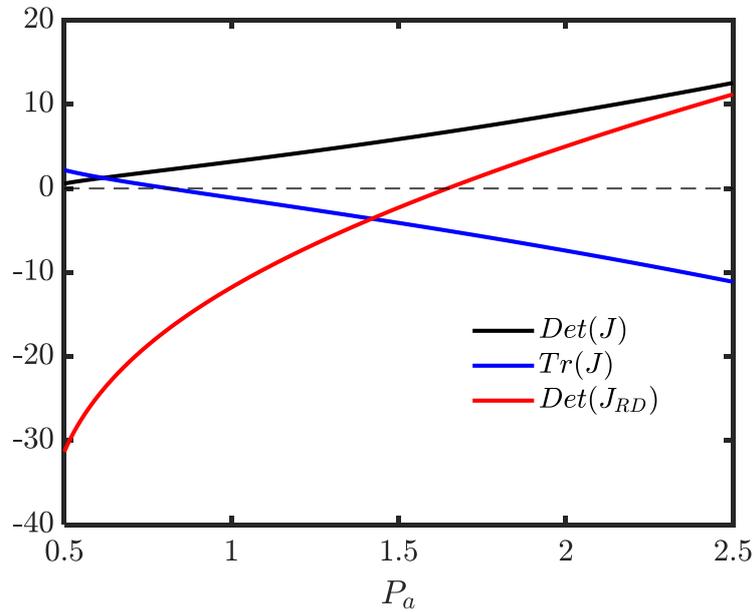

**Fig.9** The variation of $Det(J), Tr(J)$ and $Det(J_{RD})$, with parameter $P_a$ for the steady state $P_1$ for $g_b = 0.5, P_b = 1.5, d_b = 0.2, K_a = 2$ and $d = 0.1$. $Det(J) > 0, Tr(J) < 0$ and $Det(J_{RD}) < 0$ for $P_a$ lying in the range of $(0.8129 - 1.6460)$ in which Turing Patterns can arise.

We next determine the effect of diffusion on the stability of the steady states. $Det(J_{RD})$ at the steady state $P_1$ is negative for $P_a$ lying in the range of $(0.5 - 1.64)$ as shown in Fig.9. This suggests that Turing patterns can indeed emerge for these parameters. We select $P_a = 1.5$ to ensure multiple steady states exist as shown in Fig.7. We conduct extensive numerical simulations of the system to confirm the existence of Turing patterns predicted through linear stability analysis. An infinitesimally small perturbation at $P_1$, in both species concentrations of the form $0.05\sin(4x)$, is introduced as initial condition. We select the diffusivity ratio ($d = 0.1$), less than its critical value and observe that the system exhibits a spatially periodic patterns of the species shown in Fig.10. This arises from a supercritical bifurcation. On the other hand, $P_3$ is stable to small disturbances but unstable to finite amplitude disturbances given by $2\sin(4x)$. The numerical simulation for finite amplitude disturbance reveals that the system reaches the Turing pattern state which is the same as the one obtained through numerical simulation when small disturbances are imposed around $P_1$ shown in Fig.10. Hence for these parameters the spatially uniform steady state $P_3$ coexists with the Turning pattern branch from $P_1$.

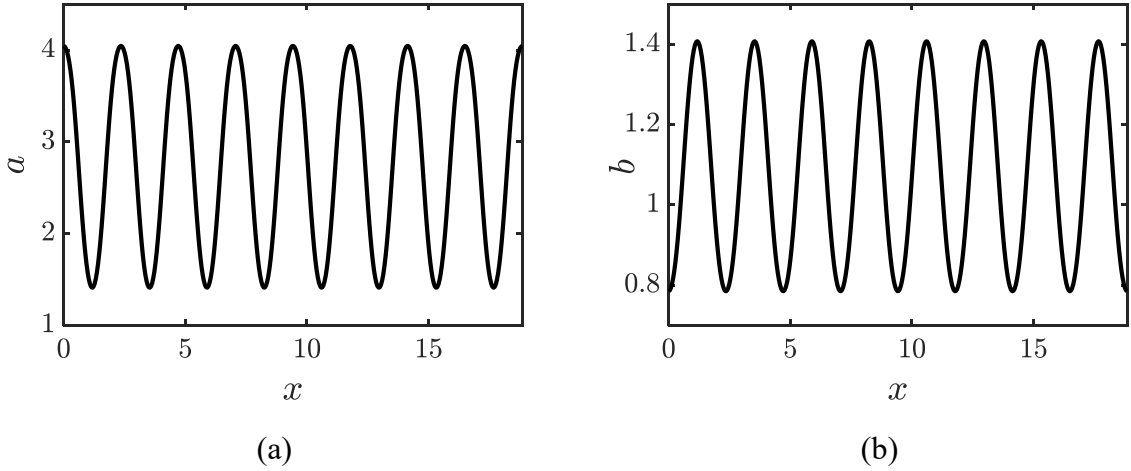

(a)  (b)

**Fig.10** Turing patterns for the parameters $g_b = 0.5, P_b = 1.5, d_b = 0.2, K_a = 2, P_a = 1.5, P_b = 1.5$ and $d = 0.1$. The initial conditions around $P_1$ are $a = 2.7845 + 0.05\sin(4x), b = 1.0773 + 0.05\sin(4x)$ while around $P_3$ are $a = 0.2154 + 2\sin(4x), b = 13.9226 + 2\sin(4x)$. Turing patterns are same for different perturbations around the steady states $P_1$ and $P_3$.

## 3.2. Turing pattern formation in 2D domain

In this section, we explore Turing pattern formation in a rectangular domain with dimensions [0, $L_x$] × [0, $L_y$], where $L_x$ and $L_y$ are the length in $x$ and $y$ directions, respectively. The approach to determine the critical bifurcation parameter value $d_c$, and its corresponding critical wave number $k_c$, is similar to that used in the 1D case (see section 3.1). Here $k_c$ refers to the composite mode in both directions and is expressed in terms of $(\phi_i, \gamma_i)$ defined as $\phi_i = n\pi/L_x$ and $\gamma_i = m\pi/L_y$.

We proceed by assuming that there is a single unstable eigenvalue, $s_1(k_c^2)$, in the instability range. For Neumann boundary conditions, we seek a solution to the linear system (19a) at $O(\varepsilon)$, as:

$$\boldsymbol{u_1} = \sum_{i=1}^{N} A_{pi}(T_1, T_2)\boldsymbol{r}\cos(\phi_i x)\cos(\gamma_i y), \tag{25}$$

where $N$ represents the multiplicity of the eigenvalue. The value of $N$ is 1 or 2 depending on whether there are one or two pairs of $(m, n)$ such that $k_c^2 = \phi_i^2 + \gamma_i^2$ and $A_{pi}$ represents slowly varying amplitudes in time and space.

*Square patterns*

A square pattern is a two-dimensional pattern where the amplitudes in both dimensions are equal, resulting in a shape with four sides of equal length and all sides meeting at right angles. The multiplicity has value $N = 1$, for rolls and square patterns [35]. We extend our WNL analysis to find the point in parameter space for square pattern. At $O(\epsilon^3)$, we derive the cubic amplitude equation which describes the evolution of amplitude ($A_{pi}$) with time (same as Stuart Landau equation (23)). The procedure to find $\sigma_1$ and $L$ for 2D case has been detailed in supplementary material Appendix B. The WNL analysis predicts that the asymptotic solution can be expressed as follows

$$\boldsymbol{u_1} = A_{p1}\boldsymbol{r}\cos(\phi_1 x)\cos(\gamma_1 y), \tag{26}$$

with $\boldsymbol{r} \in Ker(\boldsymbol{J} - k_c^2 \boldsymbol{D}_c)$.

*Case (1) Supercritical case: $L > 0$*

To obtain the parameter space for square patterns, we fix $\phi_1 = \gamma_1 = 1$ and domain size $L_x = L_y = 2\pi$. We choose a parameter combination given in caption of Fig. 11 for which the system has a steady state $(1.441, 2.297)$, and the following conditions are satisfied

$$Det\ (\boldsymbol{J}) > 0, \tag{27a}$$

$$Tr\ (\boldsymbol{J}) < 0, \tag{27b}$$

$$Det(\boldsymbol{J}_{RD}) < 0, \tag{27c}$$

$$k_c^2 = \phi_i^2 + \gamma_i^2, \tag{27d}$$

$$L > 0. \tag{27e}$$

In Fig.11(a), curve I represents the variation of $Det(\boldsymbol{J})$, with parameter $K_a$ and is positive throughout. Curve II shows the trace of the Jacobian matrix ($Tr(\boldsymbol{J})$), is negative for $K_a$ values ranging from 2.5 to 3.32. This indicates the stability of the well-mixed steady state for this range of $K_a$. Curve III shows that the determinant of the reaction-diffusion Jacobian matrix ($Det(\boldsymbol{J}_{RD})$), is negative suggesting the possibility of a Turing pattern formation. Curve IV represents the condition $k_c^2 = 2$, which is satisfied by unique pair $(m, n) = (2, 2)$ and curve V represents the variation of $L$ with $K_a$. These two curves intersect at point P, where all the conditions in (27) are satisfied. Here $L$ is positive, confirming the emergence of a square pattern through a supercritical bifurcation. Here, the Landau constant ($L = 0.142$) is positive and $K_a = 3.28$.

We perform weakly nonlinear analysis for $K_a = 3.28$ and observe that the system possesses a stable equilibrium state at which the amplitude is equal to $A_{ss} = \pm\sqrt{\sigma_1/L}$. The expressions of $L$ and $\sigma_1$ have been provided for 2D case in supplementary material Appendix B. Fig.11(b) depicts the amplitude at the steady state as a function of the control parameter $d$ in vicinity of its critical value obtained using weakly nonlinear analysis at $O(\epsilon^3)$. The amplitude increases as we go away from the critical value $d_c$ towards the left.

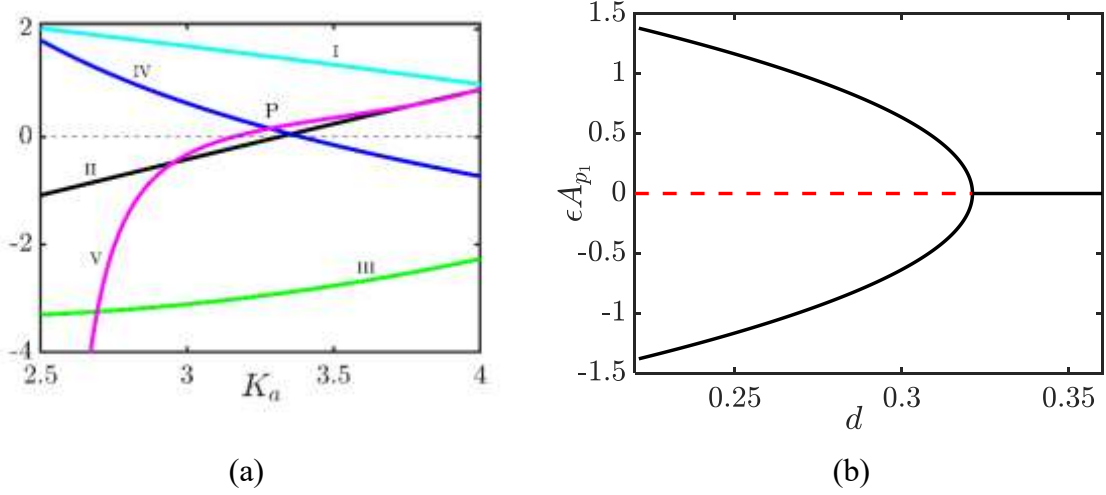

(a)            (b)

**Fig.11** (a) Determination of point P for the square pattern through supercritical bifurcation for $g_b = 0.12, P_b = 0.1, P_a = 1.8, d_b = 0.2$, and $\phi_1 = \gamma_1 = 1$. Curve I represents the variation of $Det\,(J)$, curve II represents $Tr(J)$, curve III represents $Det(J_{RD})$, curve IV represents $k_c^2 = 2$, and curve V represents $L$ , (b) The bifurcation curve describing the supercritical bifurcation for square pattern for the parameters $g_b = 0.12, P_b = 0.1, P_a = 1.8, K_a = 3.28, d_b = 0.2, \epsilon = 0.1$ and $\phi_1 = \gamma_1 = 1$. The solid lines represent the stable branches while the dashed line represents the unstable branch.

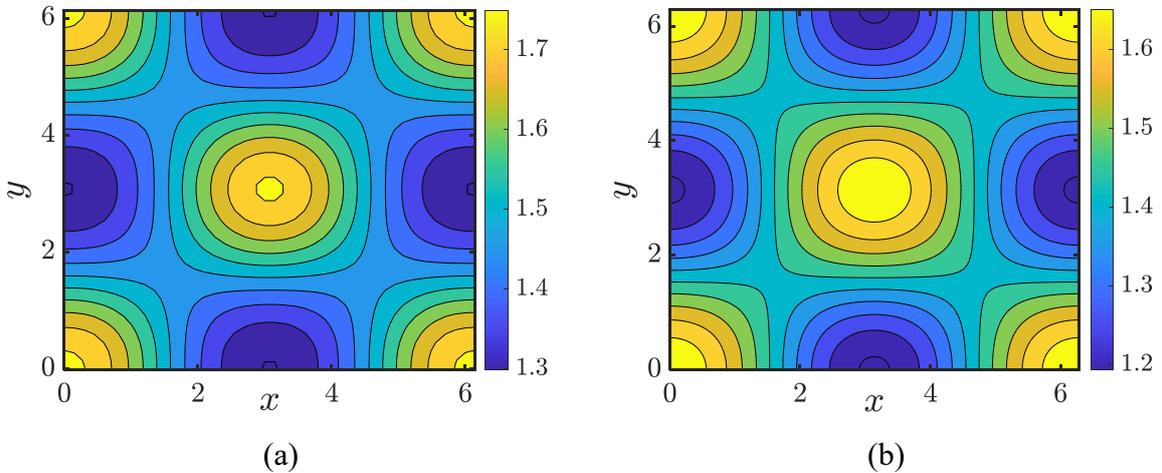

(a)            (b)

**Fig.12** Square patterns through supercritical bifurcation of species A for parameters $g_b = 0.12, P_b = 0.1, P_a = 1.8, K_a = 3.28, d_b = 0.2, d = 0.317947, \epsilon = 0.1, \phi_1 = \gamma_1 = 1$ and $L_x = L_y = 2\pi$. Contour plots from (a) Numerical solution (b) Weakly nonlinear analysis. The initial conditions for numerical simulation around the steady state are $a = a_{ss} + 0.02\sin(4x)\sin(4y)$ and $b = b_{ss} + 0.02\sin(4x)\sin(4y)$.

We also perform 2D numerical simulation to validate the amplitude predicted by weakly nonlinear approximation. We choose the operating point as P in Fig.11(a). We take domain size $L_x = L_y = 2\pi$ with time step of $\delta t = 0.1$ and spatial grid size of $\delta x = \delta y = 0.15$. The initial conditions for the numerical simulations involve a random infinitesimally small periodic perturbation around the base state. Fig.12(a) shows the contour plot of concentration of species A obtained through numerical simulations at time $t = 2000$. The variation of concentration of species $A$ in space is given by equation (28). The amplitude $\epsilon A_{ss}$ is finite for $d < d_c$ and it becomes zero for $d > d_c$. Fig.12(b) shows the concentration contour profile of species A obtained using a weakly nonlinear analysis. The amplitude obtained using weakly nonlinear analysis is in good agreement with those obtained from the numerical solution of the full system (6a-6b) as shown in Fig.12. These results validate the accuracy of framework to identify parameters for square patterns in 2D. The species vary with small amplitude confirming that the Turing patterns are obtained through supercritical bifurcation.

$$a_1 = a_{ss} + \epsilon A_{p1} \cos(\phi_1 x) \cos(\gamma_1 y). \tag{28}$$

*Case (2) Subcritical case:* $L < 0$

For the subcritical case we fix $\phi_1 = \gamma_1 = 2$ and the domain size $L_x = L_y = \pi$. The parameters (see caption in Fig.13) in this case are determined by equation (27a-27d), to ensure the condition that $L < 0$ at the steady state $(1.793, 1.066)$. In Fig.13(a), Curve I representing $(Det(J))$, is positive for all values of $K_a$. Curve II depicts that $(Tr(J))$, is negative for $K_a$ values ranging from 1 to 3.569 ensuring that the well mixed steady state is stable. Curve III shows that the determinant of Jacobian matrix $(Det(J_{RD}))$, is always negative. This suggests that Turing Patterns are possible. Curve IV represents the condition $k_c^2 = 8$, which is satisfied by unique pair $(m, n) = (2, 2)$ and curve V represents the variation of parameter $L$ with $K_a$. These two curves intersect at point P'. Here, the Landau constant ($L = -0.314$) is negative and $K_a = 1.78$. Here, we expect the emergence of a square pattern through a subcritical bifurcation. Fig.13(b) depicts the amplitude at the steady state as a function of the control parameter $(d)$ in the vicinity of critical value for $d$ obtained using weakly nonlinear analysis at $O(\epsilon^3)$. The dashed line for $d > d_c$ shows that the bifurcating branch is unstable. The numerical simulation of the full system for $d > d_c$ show spatially uniform solution which is shown by solid black line in Fig.13(b). Fig.14 shows the concentration contour plot for species A and B through numerical simulations at time $t = 2000$ for the parameters corresponding to point $P'$ in Fig.13(a). We take domain size $L_x = L_y = \pi$ with time step of $\delta t = 0.1$ and spatial grid size of $\delta x = \delta y = 0.1$. Numerical simulation of the full system for finite amplitude perturbation confirms the square patterns with a high amplitude as shown in Fig.14.

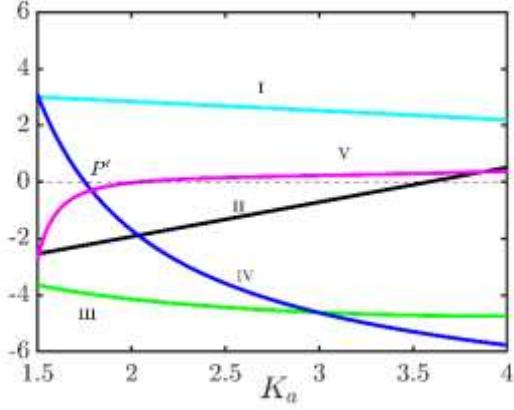 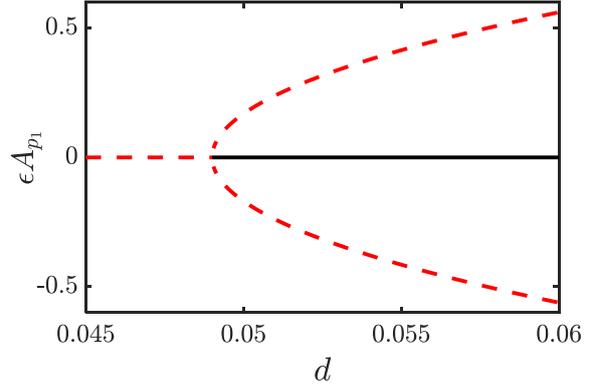

(a)            (b)

**Fig.13** (a) Determination of point $P'$ for the square pattern through subcritical bifurcation for the parameters $g_b = 0.12, P_b = 0.1, P_a = 1.8, d_b = 0.1$ and $\phi_1 = \gamma_1 = 2$. Curve I represents $Det(J)$, curve II represents $Tr(J)$, curve III represents $Det(J_{RD})$, curve IV represents $k_c^2 = 8$, and curve V represents $L$. (b) The bifurcation curve describing the subcritical bifurcation for $K_a = 1.78$. The solid black line represents the stable branch and the dashed line represents the unstable branches.

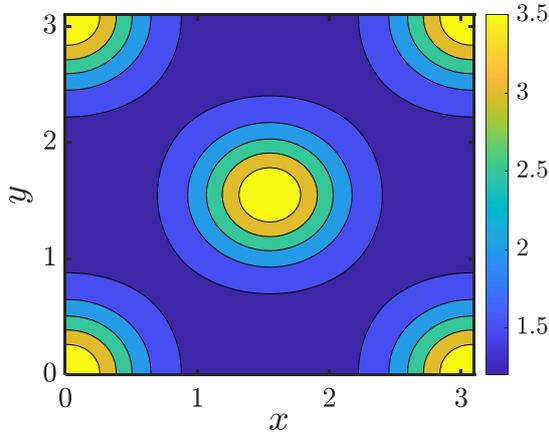 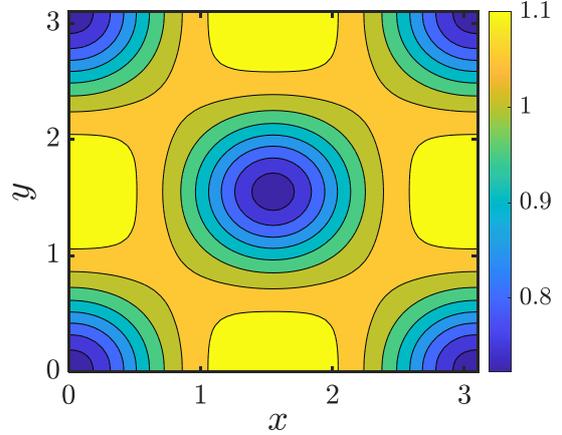

(a)            (b)

**Fig.14** Numerical simulation for square patterns through subcritical bifurcation for parameters $g_b = 0.12, P_b = 0.1, P_a = 1.8, K_a = 1.78, d_b = 0.1, d = 0.049497$ and $L_x = L_y = \pi$. (a) Concentration contour profile for species A (b) Concentration contour profile for species B. The initial conditions are $a = a_{ss} + 0.5\sin(4x)\sin(4y)$ and $b = b_{ss} + 0.5\sin(4x)\sin(4y)$.

*Hexagonal patterns*

For the same set of parameters as for supercritical square pattern, we change the domain size to $L_x = L_y = 4\pi$. For a simulation with time step of $\delta t = 0.1$ and spatial grid size of $\delta x = \delta y = 0.2$. hexagonal patterns appear when initial conditions are given as $a = a_{ss} + 0.02\sin(4x)\sin(4y)$ and $b = b_{ss} + 0.02\sin(4x)\sin(4y)$. These are shown in Figs.15(a) and 15(b). This confirms that domain size plays very important role in the type of patterns formed.

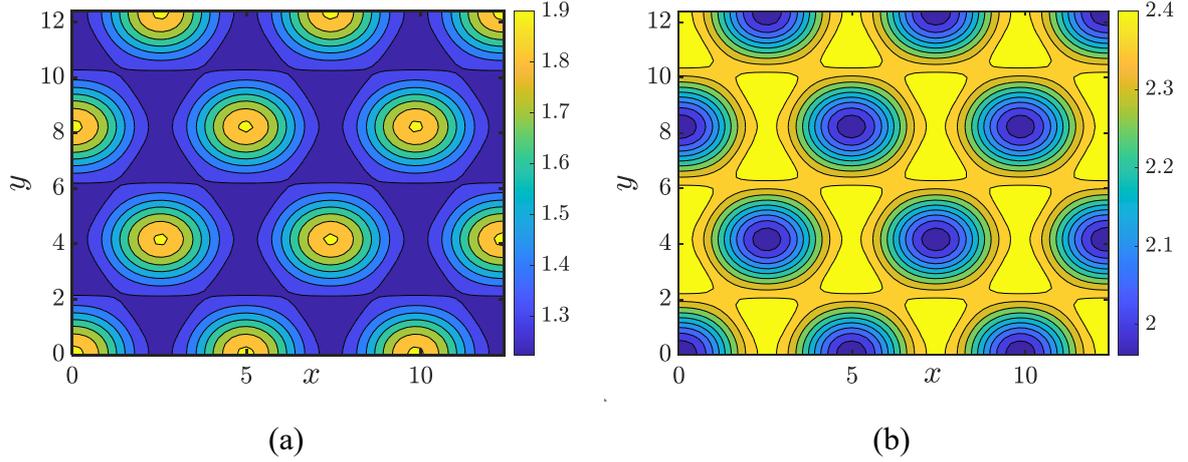

**Fig.15** Numerical simulation for hexagonal patterns for parameters $g_b = 0.12, P_b = 0.1, P_a = 1.8, K_a = 3.28, d_b = 0.2, d = 0.317947$, and $L_x = L_y = 4\pi$. (a) Concentration contour profile for species A (b) Concentration contour profile for species B. The initial conditions around the steady states are $a = a_{ss} + 0.02\sin(4x)\sin(4y)$ and $b = b_{ss} + 0.02\sin(4x)\sin(4y)$.

## 4. Conclusions

In this work, we have analysed an autocatalytic reaction diffusion system encountered in asymmetric autocatalysis. The model can be used to obtain a pure enantiomer from a mixture of enantiomer with its optical isomer. This work has focused on developing a framework to identify the region of supercritical and subcritical bifurcation for a system which has many parameters. This is achieved by a weakly nonlinear analysis of the autocatalytic reaction diffusion system near the critical value of the bifurcation parameter $(d)$.

The steady state of the spatially homogeneous system can exhibit both stationary as well as dynamic instability. We identified the regions in the bifurcation diagram where only one stable steady state exists and established that including diffusion can give rise to Turing patterns in this region. The critical surface across which Turing patterns can be formed is obtained. Using weakly nonlinear stability analysis, the regions in which the system exhibits supercritical and subcritical patterns are identified. The amplitude and the variation of the pattern in the space is validated using numerical simulations for supercritical bifurcation. In case of subcritical bifurcation, the turning / limit point is also obtained and verified numerically by extending the analysis to fifth order.

We also identified the parameters for which multiple steady states exist and Turing patterns arise by introducing diffusion through supercritical bifurcation from one of the steady states. The Turing pattern coexists with a spatially uniform steady state. The framework was extended to find the conditions and parameters for the existence of the square patterns through supercritical and subcritical bifurcation. The amplitude and shape of patterns are verified numerically. We show that the shape of the patterns is dependent on the domain size.

We have used weakly nonlinear analysis to identify the type of bifurcation. Although this helps us to obtain Turing patterns analytically, it is valid only in the vicinity of the control parameter ($d$). In case of subcritical bifurcation, the variation of species concentration is very rapid in space. Here weakly nonlinear analysis is unable to capture the large amplitude of the patterns. We have provided the theoretical and numerical analysis of Turing pattern formation in a reaction diffusion system sustaining cubic and quadratic autocatalysis. These are encountered in asymmetric autocatalysis. We believe our work will serve as a guide to determine operating condition and help experimentally validate the occurrence of Turing patterns.

Turing patterns have been hypothesized in several works as a mechanism to explain pattern formation. The results obtained in this study are relevant for ecological, chemical, and biological systems whose behavior is governed by a large number of parameters. The model used here has been studied earlier and is important in asymmetric autocatalysis [30,31]. These models have been used to obtain a purely desired enantiomer from mixture of enantiomers [28,29,37]. Additionally, the analysis in this work can be used to predict the occurrence of asymmetric drug molecules in biological systems. The mathematical framework developed in this article is useful in finding the parameters for super and subcritical bifurcation. This is a challenge since most systems have many parameters and the patterns occur in a narrow region of parameter space. Our work will help experimentalists identify and operate systems to get pattern formation.

The mathematical framework developed in this work is very general. It can be used for any nonlinear system to determine the parameters and operating conditions where the system exhibits supercritical and subcritical Turing bifurcation. This will help the experimentalist identify the narrow range of operating conditions where Turing patterns occur. The analysis can be used to explain the periodic patterns observed on animal skins like zebra, tiger.

## Acknowledgements

The author thanks Indian Institute of Technology, Madras for financial assistance.

## Supplementary material

Appendix A: Weakly nonlinear analysis in 1D.

Appendix B: Weakly nonlinear analysis in 2D.